# Lattice-Boltzmann simulations of self-assembly of a binary amphiphilic fluid into ordered bicontinuous cubic and lamellar phases

Maziar Nekovee and Peter V. Coveney

*Centre for Computational Science, Department of Chemistry, Queen Mary, University of London, Mile End Road, London E1 4NS, United Kingdom*

Periodic arrangements of two fluid media separated by interfaces are very common in liquid crystals where they most often occur as a periodic stacking of layers of fluids in one dimension. In lamellar phases of lyotropic liquid crystals, built by amphiphilic molecules in the presence of water, bilayers of amphiphiles and layers of water are alternatively stacked with flat interfaces defined by the polar heads of the amphiphiles[1]. Besides these phases which exhibit periodicity along one direction, the phase diagrams of amphiphilic systems may present other ordered phases with periodicity along two or three dimensions, curved interfaces and various other more exotic topologies.[1,2] One of the intriguing aspects of amphiphilic polymorphism is the existence of phases with long-range cubic order and bicontinuous geometry which can be traversed in any direction in both the water-rich and the surfactant-rich regions. These phases have been observed experimentally in amphiphilic-water phase diagrams close to the lamellar phase domain where the concentration of the surfactant is high.[3]

The equilibrium properties and stability of these structures have been studied in the past using macroscopic curvature models based on Helfrisch Hamiltonian, which gives the energy cost of bending a membrane-like interface[4], and in the framework of Ginzburg-Landau models[5,6]. However, the fascinating dynamics of self-assembly of such ordered structures, together with the dynamics of phase transitions, are beyond the scope of such equilibrium approaches. On the other hand, the time and length scales involved in the self-assembly of such mesophases makes fully microscopic descriptions (based on molecular dynamics) computationally prohibitive.[7] In this Communication we use our recently developed lattice-Boltzmann method[8,9] to study, on a mesoscopic level, the dynamics of self-assembly of the bicontinuous cubic phase in a binary water-surfactant system, and the transition from the lamellar structure, with flat interfaces, to a bicontinuous cubic phase, with minimum curvature.[10] Our study provides insight into how such structures emerge as a result of competing molecular interactions between water and amphiphiles and among amphiphilic molecules themselves. The present study also represents the first application of any lattice-Boltzmann model to the study of amphiphilic systems in three-dimensions.

Our lattice-Boltzmann method for simulation of ternary amphiphilic fluids is described in details in refs. 8 and 9, and with the removal of oil reduces to a model for binary water-surfactant systems. The simulations follow the time evolution of a set of single-particle distribution functions, one for water and one for amphiphilic molecules. The distribution functions evolve according to a set of coupled Boltzmann equations which are discretized in time, space and velocity. The coupling between fluid components is achieved by introducing self-consistently generated pair-wise forces between water and amphiphiles and among amphiphiles themselves. The presence of these forces results in off-diagonal and cross-component collision terms in the lattice-Boltzmann equations.[9] Amphiphilic molecules have polar headgroups that are hydrophilic, and hydrocarbon tails that are hydrophobic, and so there is a strong energetic preference for these to create water-rich and amphiphilic-rich regions such that the hydrophilic headgroups are in contact with water while hydrophobic tails are shielded from water. Moreover, the presence of these orientable particles on the interfaces gives it a bending or curvature energy. In our approach these important features are taken into account by assuming that amphiphilic molecules possess an orientational degree of freedom which can vary continuously in time and space according to a Boltzmann-type equation. This is a crucial ingredient of our approach and, as was shown in refs. 8 and 9, allows our model to describe correctly the essential phenomenology of amphiphilic systems. The intermolecular forces act only between nearest-neighbors and their strengths are controlled by a set of couplings $g_{ws}$ and $g_{ss}$ for water-surfactant and surfactant-surfactant interactions, respectively. The relaxation times $\tau_w$ and $\tau_s$ determine the kinematic viscosities of each component.[8] Finally, the inverse temperature $\beta$ mimics the tendency of temperature to disorder the orientational preference of amphiphiles.

Our simulations were performed on a $64^3$ cubic cell, satisfying periodic boundary conditions. The force couplings, relaxation times and inverse temperatures used were $g_{ws} = -0.01$, $g_{ss} = 0.0015$, $\tau_w = \tau_s = 1$ and $\beta = 10$. The average concentrations of water and surfactant were set at $0.2$ and $0.1$, respectively. These choices of parameters were made after a limited search in the parameter space of the model, taking into account that, due to their dipolar nature, the amphiphile-amphiphile interactions are intrinsically very strong, hence $g_{ss}$ should be chosen small such that these interactions do not completely dominate. Furthermore the inverse temperature $\beta$ should be large enough to ensure that intermolecular interactions are able to overcome the disordering tendency of the temperature. To check finite-size effects we performed an additional set of simulations on a $32^3$ system and found essentially the same phenomenology as in the case of the $64^3$ system.

In Fig. 1 we show snapshots of our simulations of self-assembly within the $64^3$ system, where the mid-surfaces of surfactant density are shown (the water density shows a similar behavior). It can be seen that starting from a homogenized mixture of water and surfactant (equivalent to a thermal quench), the system self-assembles into a triply periodic bicontinuous structure of water-rich and surfactant-rich regions which fill up the whole space. Pictorially, it can be seen that the mid-surface of the cubic structure corresponds to Schwarz's P surface, one of the triply periodic minimal surfaces (TPMS).[11] Mathematically TPMS are defined as surfaces with mean zero curvature $C_1 + C_2 = 0$, where $C_1$ and $C_2$ are the principal curvatures of a film embedded in three-dimensions. The lamellar structure corresponds to $C_1 = C_2 = 0$ while the TPMS are obtained when $C_1 = -C_2$. It was shown by Luzatti[12] and co-workers that the mid-surfaces of the lipid bilayers are very close to cubic minimal surfaces and they have been found in a variety of other real structures such as silica mesophases, lyotropic colloids and detergent films.[11]

In macroscopic curvature models of amphiphilic-water systems, the formation of bicontinuous cubic phases is explained as resulting from the incompatibility, in the lamellar phase, of maintaining a constant distance between the interfaces and the existence of curvature of that interface: a geometrical frustration which is resolved through the formation of interfaces having the TPMS structure.[10] In our model such behavior is an emergent property of the underlying mesoscale dynamics resulting from intermolecular forces: the amphiphile-water forces favor the formation of flat interfaces between water-rich and amphiphile-rich regions and tend to align the amphiphilic molecules normal to the interfaces, with their hydrophilic heads pointing towards the water-rich regions. On the other hand, the dipolar amphiphile-amphiphile interactions can favor attraction or repulsion between two amphiphilic heads. Since amphiphilic molecules mainly reside at the interface, it is this interaction which gives the system an additional curvature energy. To test this idea, we performed additional simulations in which we kept the water-amphiphile coupling the same but switched off all the amphiphile-amphiphile interactions by setting $g_{ss} = 0$. As can be seen from Fig. 2, after a transition state in which the cubic and the lamellar phases coexist, the system self-assembles into a stable lamellar structure, with essentially zero curvature, and the transition to a bicontinuous cubic structure does not occur. The orientation of amphiphilic molecules is not shown here but we have confirmed that, starting from initial random orientations, these molecules orient themselves in the direction normal to the lamellar structures, with their hydrophilic heads pointing towards the water-rich regions.

Our simulations thus provide evidence, on a mesoscopic level, that interactions between amphiphilic molecules is indeed the driving force that destroys the lamellar structure in favor of the bicontinuous cubic phase.

**ACKNOWLEDGMENT**. This work was partially supported by EPSRC under Grant No GR/M56234. Our simulations were performed on the Origin 2000 at the New England Supercomputing Center of Boston University. We thank Dr Bruce M. Boghosian for providing us with access to this facility.

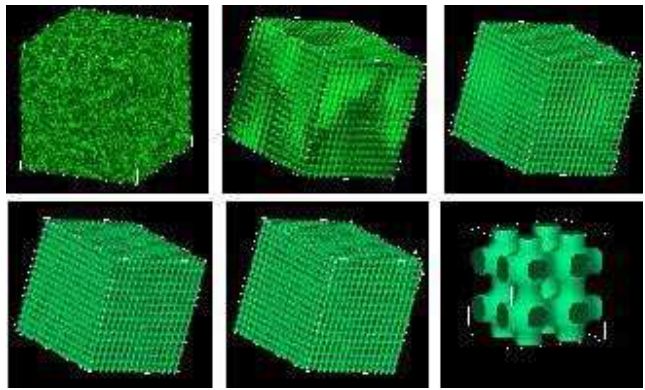

**Figure 1.** Self-assembly of a bicontinuous cubic structure in a binary water-surfactant system. The system size is $64^3$. The water and surfactant concentrations are 0.2 and 0.1, respectively. The coupling parameters are $g_{ws}$=-0.01 and $g_{ss}$=0.0015. Surfactant mid-surfaces are shown from left to right and top to bottom at time steps 0,400,1200,2000,4800. The last snapshot shows 2 periods in the x, y and z directions for the cubic structure at time step 4800.

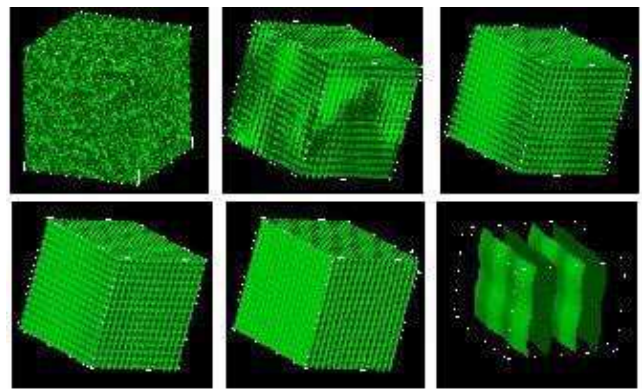

**Figure 2.** Self-assembly of a lamellar structure in a binary water-surfactant system, when the interactions among amphiphilic molecules is switched off. The water and surfactant concentrations as well as the water-surfactant coupling are kept the same as in Fig. 1. The same time steps are shown as in Fig. 1. The last snapshot shows the same volume as the last snapshot in Fig. 1.